\documentclass[journal]{IEEEtran}

\usepackage{graphicx}
\usepackage{amsmath}
\usepackage{amssymb}
\usepackage{caption}
\usepackage{subcaption}
\usepackage{float}
\usepackage{multirow}
\usepackage{csquotes}
\usepackage{subfig}
\usepackage[labelfont=bf,labelfont=normalsize,justification=raggedright,,singlelinecheck=off]{caption}
\begin{document}
\title{A 700$\mu$W 1GS/s 4-bit Folding-Flash ADC in 65nm CMOS for Wideband Wireless Communications}

\author{Bayan Nasri, Sunit P. Sebastian, Kae-Dyi You, RamKumar RanjithKumar, Davood Shahrjerdi

Electrical and Computer Engineering, New York University, Brooklyn, NY 11201

Email: davood@nyu.edu

}

\maketitle
\pagestyle{empty}
\thispagestyle{empty}

\begin{abstract}
We present the design of a low-power 4-bit 1GS/s folding-flash ADC with a folding factor of two. The design of a new unbalanced double-tail dynamic comparator affords an ultra-low power operation and a high dynamic range. Unlike the conventional approaches, this design uses a fully matched input stage, an unbalanced latch stage, and a two-clock operation scheme. A combination of these features yields significant reduction of the kick-back noise, while allowing the design flexibility for adjusting the trip points of the comparators. As a result, the ADC achieves SNDR of 22.3 dB at 100MHz and 21.8 dB at 500MHz (i.e. the Nyquist frequency). The maximum INL and DNL are about 0.2 LSB. The converter consumes about 700$\mu$W from a 1-V supply yielding a figure of merit of 65fJ/conversion step. These attributes make the proposed folding-flash ADC attractive for the next-generation wireless applications.

\end{abstract}

\section{Introduction}

Next-generation 5G wireless communications would use millimeter wave bands between 30 and 300GHz \cite{Barati}. The use of such wide spectrum allows the implementation of transceivers with high-dimensional antenna arrays for analog or digital beamforming. For many applications, however, the overall power consumption of the transceiver would be a key design parameter. In analog beamforming, the incoming signals from the antenna array are combined in the analog domain and processed by a single pair of ADC. In contrast, the digital beamforming uses a pair of ADC for each antenna element. The architecture of the digital beamforming is more flexible than the analog beamforming, thus making the digital beamforming more popular for the implementation of cellular transceivers \cite{Barati}. However, for a transceiver with large antenna array, the high number of ADCs might lead to significant increase in the power consumption. A recent report by Orhan {\it{et. al}} shows the possibility of reducing the overall power consumption of a fully digital transceiver without compromising its performance by reducing the bit resolution of the high-speed ADCs \cite{O.Orhanl}. 

Previous works have shown high-speed flash ADCs with low-bit resolution for applications in wideband transceivers \cite{Van_der_Plas,Nuzzo}. Although conventional flash ADCs have high speed, their power consumption is high because they require $2^N -1$ comparators for an N-bit conversion. The application of signal folding technique in flash ADCs can reduce the number of comparators while maintaining the high conversion rates, thereby giving rise to significant improvement of key design parameters including the power consumption, the kickback noise, and the chip area \cite{Verbruggen,Amico}.  

In this work, we introduce a low-power 4-bit 1GS/s folding-flash ADC with folding factor of two. We propose an unbalanced comparator, which significantly improves the power consumption and the kick-back noise of the ADC. The ADC consumes as low as 700$\mu$W from a 1V supply, giving a figure of merit (FoM) of 65fJ/conversion step. The paper structure is as follows: section II describes the system architecture, section III describes the key design considerations of the comparator, and section IV presents the simulation results.

\section{System Architecture}

\begin{figure}[h!]
\centering
\includegraphics[width=3.5in]{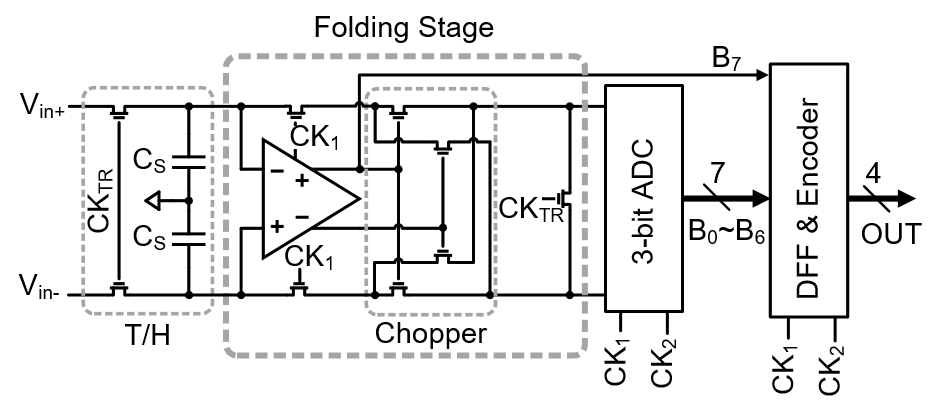}
\caption{Architecture of the proposed folding-flash ADC.}
\label{fig_sim}
\end{figure}

Fig.\ref{fig_sim} shows the architecture of the proposed folding-flash ADC. The ADC comprises a track-and-hold (T/H) circuit, a 1-bit folding stage, a 3-bit flash ADC, and a digital encoder. Two important design parameters in high-speed folding-flash ADCs are the linearity of the folding stage and the kick-back noise, where the kick-back noise arises from the high frequency switching in comparators. To increase the linearity, the sampling capacitor ($C_s$) should be larger than the total parasitic capacitance at the input nodes of the comparator \cite{Verbruggen}. A sufficiently large $C_s$ can also reduce the kick-back noise, generated primarily by the comparators in the 3-bit ADC \cite{Amico}. The optimal value of $C_s$ in our design is 500fF. 

\begin{figure}[h!]
\centering
\begin{subfigure}[t]{\linewidth}
        \centering
        \caption{}\label{timing}
        \includegraphics[width=3.0in]{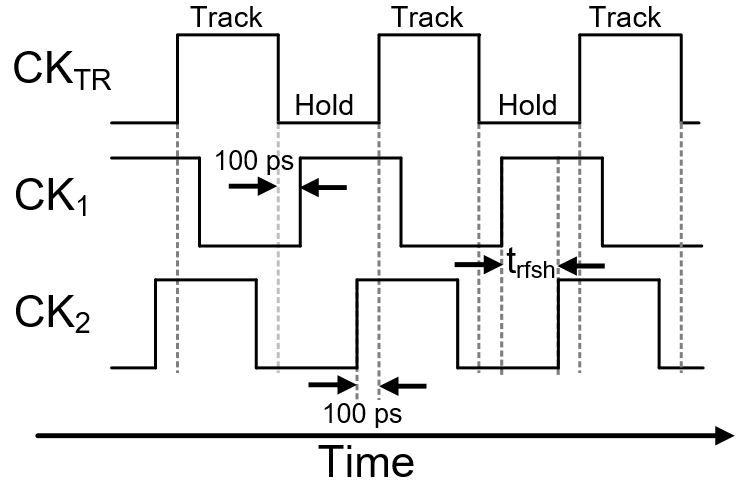}
    \end{subfigure} \\
    
\begin{subfigure}[t]{\linewidth}
        \centering
        \caption{}\label{inv}
        \includegraphics[width=2.8in]{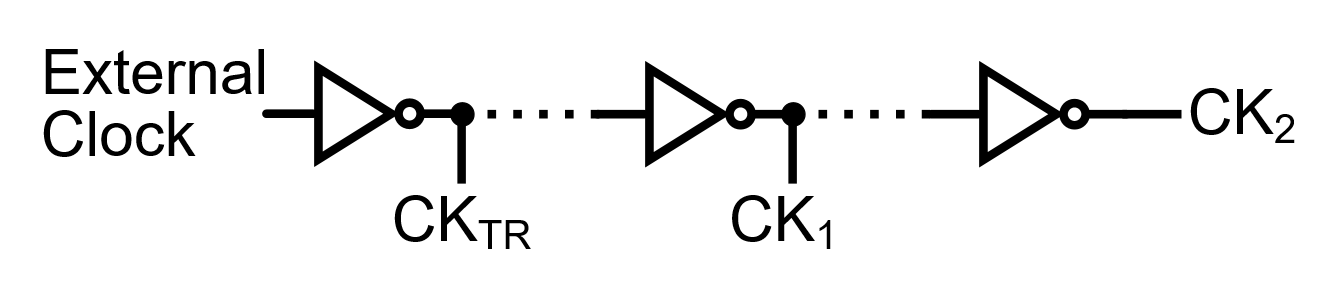}
    \end{subfigure} \\

\caption{(a) Timing diagram of the folding-flash ADC (b) Inverter delay chain to produce different clocks for ADC.}

\end{figure}

Fig.\ref{timing} illustrates the timing diagram of the clock signals. All clock waveforms are generated by feeding an external clock to an inverter chain to produce the desired timing of these signals, shown in Fig.\ref{inv}. To improve the accuracy of the ADC, $CK_1$ and $CK_2$ signals are generally out of phase with respect to the $CK_{TR}$ signal. In our design, $CK_1$ occurs 100ps after the hold phase to accommodate for the settling time of $C_s$, while $CK_2$ takes place 100ps before the next track cycle to account for the 3-bit ADC decision time.

The chopper circuit shares the charge on the sampling capacitor $C_s$  with the parasitic input capacitance of the 3-bit flash ADC during $CK_1$. The parasitic input capacitance of the 3-bit ADC is reset to zero during the tracking phase to remove the residual charge from the previous sample. In our design, the comparators in the 3-bit ADC are unbalanced with built-in references to quantize the input signal. The output signals of the 3-bit flash ADC are in the form of a thermometric code. The code is then passed to a digital encoder, which incorporates a first-order bubble correction for producing a more accurate gray code.

\section{Circuit design}
\subsection{Conventional double-tail comparator}

\begin{figure}[h!]
\centering
\begin{subfigure}[t]{\linewidth}
        \centering
        \caption{}\label{CMP_con}
        \includegraphics[width=3.0in]{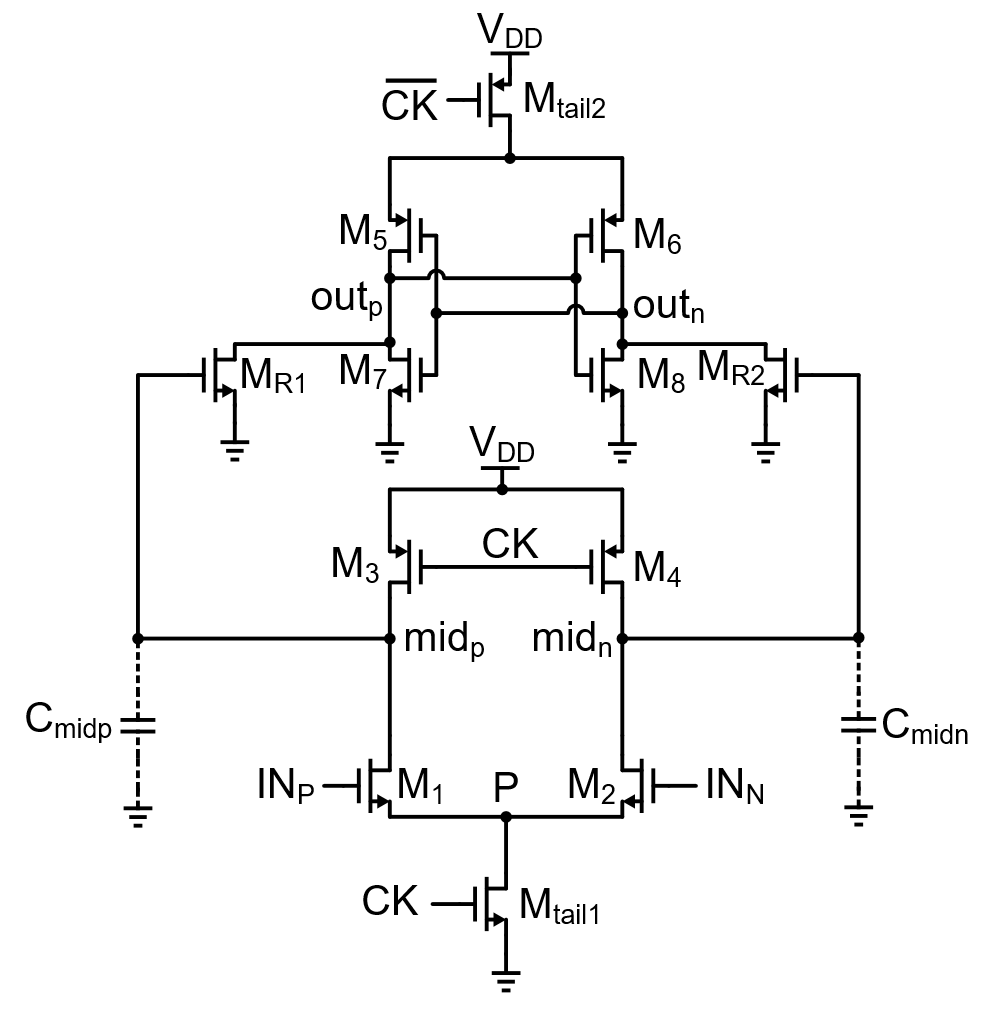}
    \end{subfigure} \\
    
\begin{subfigure}[t]{\linewidth}
        \centering
        \caption{}\label{CMP}
        \includegraphics[width=3.0in]{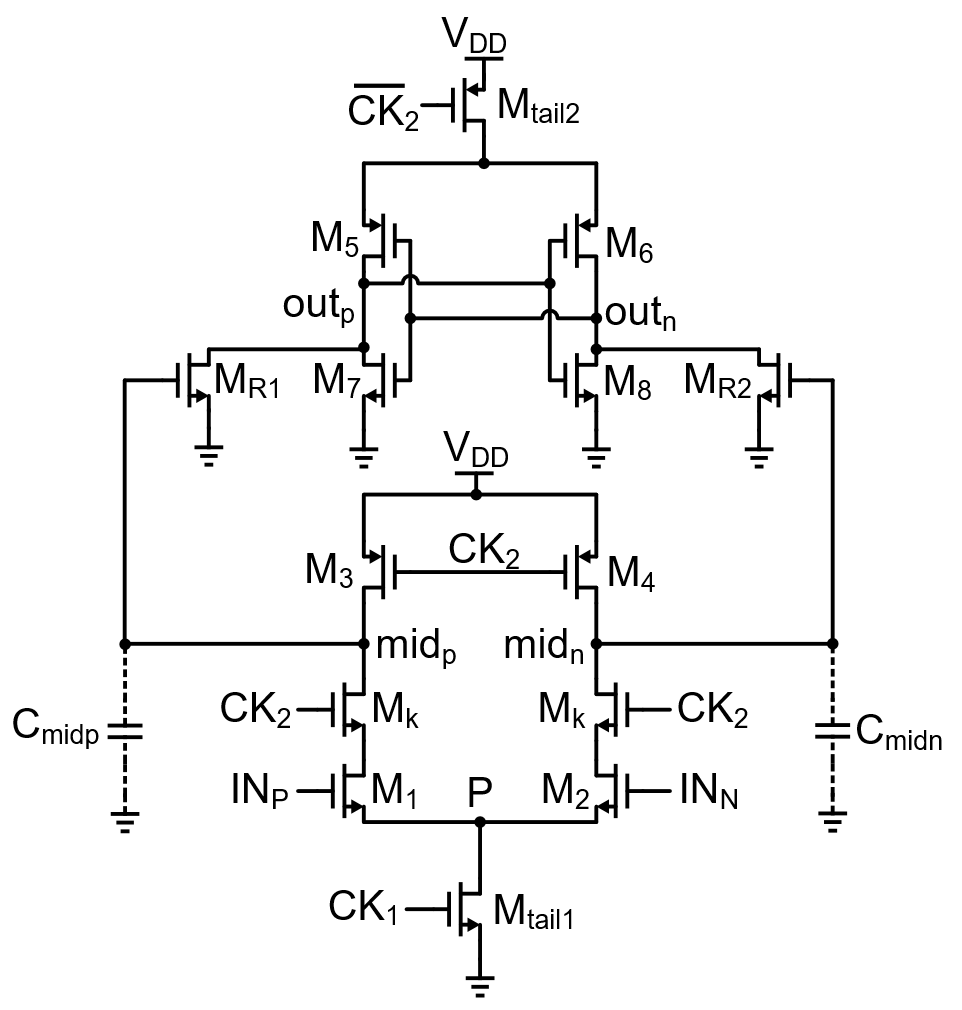}
    \end{subfigure} \\

\caption{Schematic of (a) conventional double-tail comparator and (b) proposed double-tail comparator.}

\end{figure}

Fig.\ref{CMP_con} shows the transistor-level architecture of a double-tail comparator. The double-tail comparator is commonly used in data converters due to its high speed, low offset, and low static power consumption \cite{D.Schinkel}. 

To eliminate the conventional resistor ladder in the flash ADC architecture, it is desirable to implement built-in references. This is typically done by introducing an intentional offset at the input of the comparator. There are different methods for implementing this offset. One approach involves using different size input transistors \cite{Van_der_Plas,Nuzzo} to make their transconductance different from one another. This, however, makes the input capacitance highly imbalanced, which can result in unpredictable kick-back noise and degrade the linearity. 

Another approach to program the offset is by varying the capacitive load difference at the $mid_p$ and $mid_n$ nodes of the circuit ($C_{diff}=C_{midp}-C_{midn}$) and keeping the input $M_1$ and $M_2$ transistors balanced. The shift in the trip point (i.e. $V_{offset}$) is given by the following expression \cite{Nikoozadeh}:

\begin{equation} \label{eq:1}
V_{offset}=\frac{I_{D}}{g_{m1,2}}.\frac{C_{diff}}{C_{sum}}=\frac{V_{ov1,2}}{2}.\frac{C_{diff}}{C_{sum}}
\end{equation}

where $C_{sum}$ is the total load capacitance in the balanced case, $I_D$, $g_{m1,2}$ and $V_{ov1,2}$ are the drive current, transconductance, and overdrive voltage of the input pair in saturation region. To implement the capacitive load difference, Verbruggen {\it{et al.}} added an MOS capacitor at the $mid_p$ node of the circuit \cite{Verbruggen}. This modification changes the slew rate of the input transistors, and the resulting regenerative action of the cross-coupled inverters for implementing the built-in reference. The potential drawbacks of this approach are the reduced linearity and larger comparator size. Alternatively, D'Amico {\it{et al.}} has implemented the offset by mismatching the size of the $M_3$ and $M_4$ transistors \cite{Amico}. One shortcoming of this implementation is the difference in the amount of the charge injection at the $mid_p$ and $mid_n$ nodes during the discharge phase, which can potentially give rise to an erroneous decision by the comparator.    

\subsection{Proposed double-tail comparator}

To overcome the aforementioned shortcomings of the previous implementations, we introduce a new double-tail comparator, shown in Fig.\ref{CMP}. The new features of our design include: (1) implementation of the built-in reference by choosing different size reset transistors $M_{R1}$ and $M_{R2}$, (2) reduction of the kick-back noise by adding the intermediate $M_k$ transistors and implementing a two-clock operation. It should be noted that the use of a two-clock operation without adding the $M_k$ transistors results in the increase of the power consumption. We describe these features in the subsequent sections. 

\subsubsection{\textbf{Implementation of built-in reference}} The 3-bit ADC consists of 7 comparators with different trip points. For each comparator, we implement the offset by mismatching the size of the reset transistors. To explain the mechanism for creating the built-in offset in our circuit, we refer to Fig.\ref{CMP_con}. Transistors $M_1$, $M_2$, $M_3$, $M_4$, $M_{R1}$ and $M_{R2}$ contribute to the total parasitic capacitance ($C_{sum}$) at the $mid_p$ and $mid_n$ nodes. Assuming that $M_1$, $M_2$, $M_3$, and $M_4$ transistors are fully matched, only the reset transistors contribute to the difference in the capacitive loads at those nodes:

\begin{equation}\label{eq:2}
C_{diff}=C_{gsR1}+C_{gdR1}(1+A_{cc})-[C_{gsR2}+C_{gdR2}(1+A_{cc})]
\end{equation}

where $A_{cc}$ is the gain of the cross-coupled inverter. Further, given the large gain of the cross-coupled inverter, we ignore the contribution of the $M_1$, $M_2$, $M_3$, $M_4$ transistors to $C_{sum}$. We can therefore express $C_{sum}$ as:

\begin{equation}\label{eq:3}
C_{sum}=C_{gsR1}+C_{gdR1}(1+A_{cc})+[C_{gsR2}+C_{gdR2}(1+A_{cc})]
\end{equation}

Since the reset transistors are operating in the triode region, the parasitic capacitance of the transistors is given by:

\begin{equation}\label{eq:4}
C_{gs}=C_{gd}=C_{ox}\times W\times L
\end{equation}

Combining the equations \ref{eq:2}, \ref{eq:3}, and \ref{eq:4}, and also assuming that the reset transistors have similar gate length {\it{L}}, we re-write the equation \ref{eq:1} as:

\begin{equation}\label{eq:7}
V_{offset}=\frac{V_{ov1}}{2}\frac{W_{R1}-W_{R2}}{W_{R1}+W_{R2}}
\end{equation}

We use this simplified model to estimate the size of the reset transistors. Although the offset can be easily implemented by mismatching  the size of the reset transistors, it is critical to match the capacitive load at the $out_p$ and $out_n$ nodes. Otherwise, this might result in a significantly large, undesirable offset \cite{Nikoozadeh}. Therefore, to match the load at these nodes, we connect the output nodes of each comparator to a buffer inverter.

\subsubsection{\textbf{Reduction of kick-back noise}} The kick-back noise occurs due to high-frequency voltage swings across the input transistors of a comparator. The cumulative kick-back noise of all comparators in the 3-bit ADC can be large enough to corrupt the sampled signal. Therefore, it is essential to reduce the kick-back noise of each comparator. We now proceed to explain the origin of the kick-back noise and strategies for mitigating it. 

In a double-tail comparator, the decision phase starts when the $CK$ signal transitions from the low state to the high state. At this time, the $mid_p$ and $mid_n$ nodes begin to discharge into the ground. This subsequently lowers the drain-source voltage of the input transistors and pushes them from the saturation region into the triode region. The change in the operating region of the input transistors creates a kick-back charge that results in a noise (i.e. kick-back noise) at the input nodes of the comparator. Therefore, the kick-back noise corrupts the sampled signal in a single-clock operation scheme. To mitigate this problem, we used a two-clock scheme for operating the 3-bit ADC comparators, shown in Fig.\ref{CMP}. In our scheme, there is enough time to refresh the input of the 3-bit ADC during $t_{rfsh}$ before the decision phase, thereby mitigating the effect of the kick-back noise. However, using the two-clock operation for a conventional double-tail comparator will significantly increase the static power consumption due to the direct path between the supply voltage $V_{DD}$ and the ground during $t_{rfsh}$. To alleviate this problem, we have added the intermediate $M_K$ transistors. The size of these transistors influence the kick-back noise. Therefore, we optimized the size of these transistors to diminish the kick-back noise during the decision phase.

\begin{figure}[h!]
\centering
\includegraphics[width=3.3in]{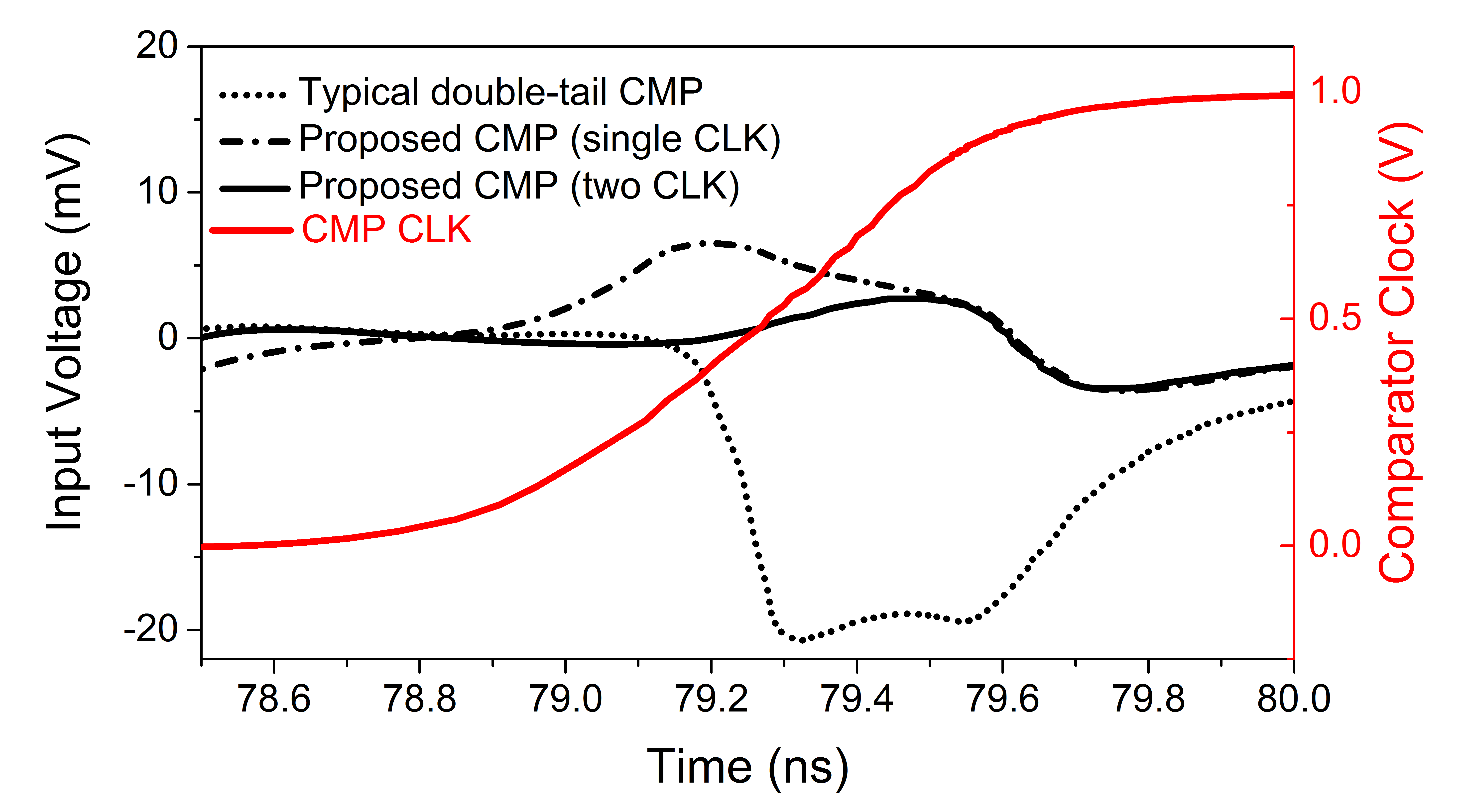}
\caption{Effect of the kick-back noise on the sampled signal at the input node of the proposed and the conventional double-tail comparators.}
\label{Kickback}
\end{figure}

We also used the proposed architecture shown in Fig.\ref{CMP} for implementing the front-end folding comparator. Unlike the comparators in the 3-bit ADC, this comparator is fully balanced (no mismatch between $M_{R1}$ and $M_{R2}$) and uses a single-clock operation scheme. Finally, we have verified the possibility to calibrate the ADC against any process variations using the bulk voltage trimming \cite{Yao}.

\section{Simulation results}

\begin{figure}[t!]
\centering
\includegraphics[width=3.0in]{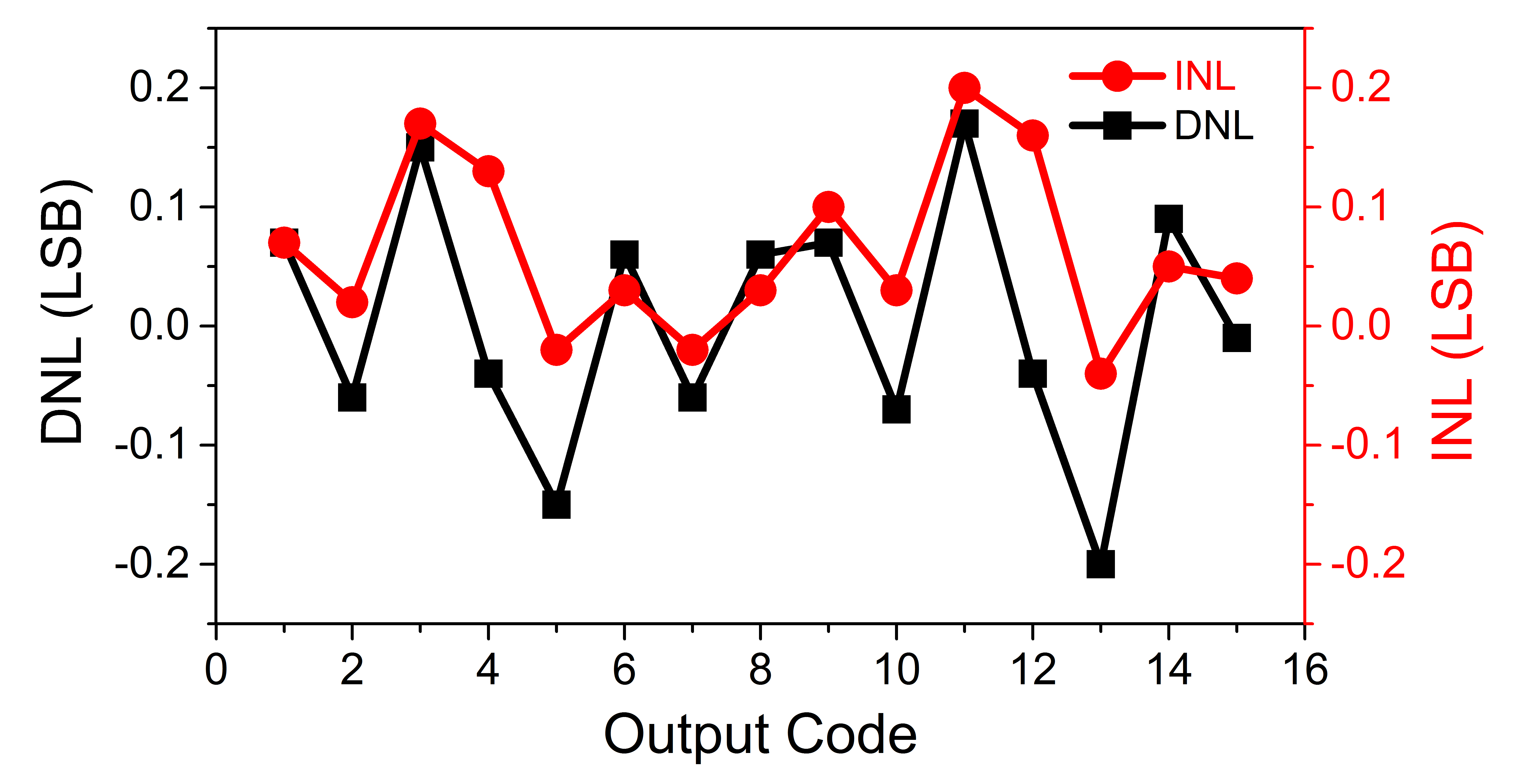}
\caption{DNL and INL for different output code words.}
\label{DN_INL}
\end{figure}

The proposed ADC was designed in a standard 65nm CMOS process with a supply voltage of 1V to operate at 1GS/s. A differential signal of $500mV_{p-p}$ is given to the input of the ADC. Fig.\ref{Kickback} shows the effect of the kick-back noise at 1GS/s. According to our simulation results, the conventional double-tail comparator exhibits a kick-back noise of around 20mV (0.64LSB) while this value is about 3mV (0.1LSB) for the unbalanced comparator with a double-clock operation scheme. Fig.\ref{DN_INL} shows the simulation results for the system linearity, where the maximum DNL and INL are less than 0.2LSB.

\begin{figure}[h!]
\centering
\includegraphics[width=3.3in]{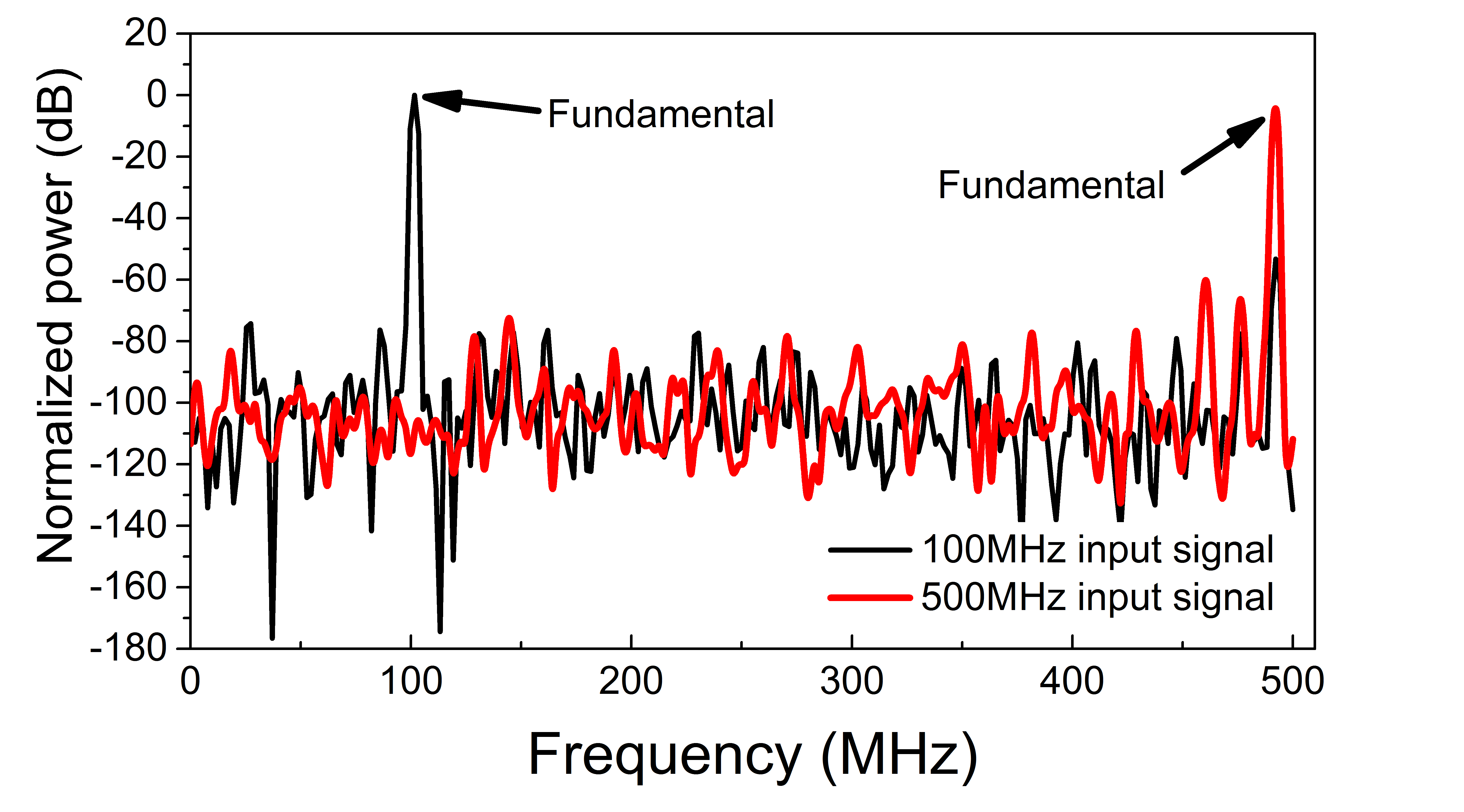}
\caption{Power spectral density of 100MHz and 500MHz input signals sampled at 1GS/s.}
\label{PSD}
\end{figure}

The FFT plots for the input frequencies of 100MHz and 500MHz are shown in Fig.\ref{PSD}. The SNDR and ENOB are 22.3dB and 3.42 bits at 100MHz while they are 21.8dB and 3.34 bits at 500MHz. Table\ref{tabel.2} summarizes the performance of the ADC. The ADC consumes about 700$\mu$W of which the T/H circuit, the comparators, the clock generator, and the encoder consume about 10\%, 25\%, 45\% and 20\%, respectively. Figure of Merit (FoM) of the system was evaluated using the following expression \cite{Murmann}:

\begin{equation}\label{eq:11}
FoM=\frac{power}{2^{ENOB}\times f_{sample}}
\end{equation}
We deduced the FoM of the ADC to be 65fJ/conversion step. Table\ref{table.3} summarizes the comparison of our folding flash ADC with other high-speed ADCs. 

\begin{table}[h!]
\centering
\caption{Performance summary}
\label{tabel.2}
\begin{tabular}{|c|c|}
\hline
Technology     & 65nm CMOS                                                       \\ \hline
Supply voltage & 1 V                                                             \\ \hline
Sampling rate  & 1GS/s                                                           \\ \hline
Number of Bits & 4                                                               \\ \hline
Input Swing    & \begin{tabular}[c]{@{}c@{}}$500mV_{p-p}$\\ \end{tabular} \\ \hline
ENOB at 100MHz  & 3.42 bits                                                                 \\ \hline
ENOB at 500MHz  & 3.34 bits                                                                 \\ \hline
INL/DNL        & 0.2 LSB                                                         \\ \hline
SNDR at 100MHz  & 22.3 dB                                                                 \\ \hline
SNDR at 500MHz  & 21.8 dB                                                                \\ \hline
Power          & 700$\mu$W                                                           \\ \hline
\end{tabular}
\end{table}

\begin{table}[h!]
\centering
\caption{Comparison of the proposed folding-flash ADC with other high-speed ADCs}
\label{table.3}
\begin{tabular}{|c|c|c|c|c|c|c|}
\hline
Ref. & Architecture & \begin{tabular}[c]{@{}c@{}}Power\\ (mW)\end{tabular} & \begin{tabular}[c]{@{}c@{}}Fs\\  (GHz)\end{tabular} & \begin{tabular}[c]{@{}c@{}}Res.\\ (Bits)\end{tabular} & \begin{tabular}[c]{@{}c@{}}SNDR\\ (dB)\end{tabular} & \begin{tabular}[c]{@{}c@{}}FoM\\ (fJ/conv.)\end{tabular} \\ \hline

{[}3{]} & Flash & 2.5 & 1.25 & 4 & 23.8 & 160 \\ \hline

{[}5{]} & \begin{tabular}[c]{@{}c@{}}Folding \\ Flash\end{tabular} & 2.2 & 1.75 & 5 & 28.5 & 50 \\ \hline
{[}6{]} & \begin{tabular}[c]{@{}c@{}}Folding \& \\ Int. Flash\end{tabular} & 7.65 & 1 & 5 & 27.4 & 390 \\ \hline
{[}8{]} & Delay line & 1 & 1 & 4 & 21.3 & 126 \\ \hline
\begin{tabular}[c]{@{}c@{}}This \\ Work\end{tabular} & \begin{tabular}[c]{@{}c@{}}Folding \\ Flash\end{tabular} & 0.7 & 1 & 4 & 22.3  & 65 \\ \hline
\end{tabular}
\end{table}

\vspace{-0.3em}

\section*{Conclusion}

We introduced an unbalanced comparator architecture for realizing ultra-low power high-speed ADCs with low-bit resolution. We designed a 4-bit 1GS/s ADC in 65nm CMOS, which consumes 700$\mu$W. This translates into an FoM of 65fJ/conversion step. These attractive specifications of our ADC make it promising for emerging applications in wideband communications such as fully digital transceivers~\cite{O.Orhanl}.

\section{ACKNOWLEDGMENTS} 

The authors acknowledge Prof. Sundeep Rangan for helpful discussions. This work is supported in part by NYU WIRELESS Industrial Affiliates program.

\end{document}